\documentclass{aa}
\usepackage{graphics}

\usepackage{graphicx}

\newcommand{\mic}{$\mu$m}

\newcommand{\msun}{\,M$_{\odot}$}

\begin{document}


\title{Determination of limits on disc masses around six pulsars\\ 
at 15 and 90~\mic\ 
        \thanks{Based on observations with ISO,
        an ESA project with instruments funded by ESA Member States
        (especially the PI countries: France, Germany, the Netherlands
        and the United Kingdom) and with the participation of ISAS and NASA.}}
\author{Lydie Koch-Miramond\inst{1}
\and Martin Haas\inst{2}
\and Eric Pantin\inst{1}
\and Philipp Podsiadlowski\inst{3}\and \\
 Tim Naylor\inst{4}
\and Marc Sauvage\inst{1}}
\institute{DAPNIA/Service d'Astrophysique, CEA/Saclay, 91191 
Gif-sur-Yvette Cedex, France
\and Max Planck Institut f\"ur Astronomie, D-69117 Heidelberg, K\"onigstuhl 17 Germany
\and Nuclear and Astrophysics Laboratory, University of Oxford, Oxford, OX1 3RH, UK
\and Astrophysics group, School of Physics, University of Exeter, EX4 4QL, UK}
\titlerunning{Mass Limits around six Pulsars}
\offprints{Lydie Koch-Miramond (e-mail: lkochmiramond@cea.fr)}
\date{Received October 18, 2001; revised March 7, 2002 / Accepted March 13, 2002}
\date{\today}

\abstract{
We have searched for evidence of emission at 15~$\mu$m with ISOCAM and
at 90~$\mu$m with ISOPHOT from dust orbiting six nearby pulsars, both in binaries 
and in isolation, located at distances between about 100 to 1000 pc. No emission was 
detected at any of the pulsar positions, and 
for the nearest pulsar J0108$-$1431 the 3 $\sigma$ upper limits on the flux density 
is about 66 mJy at 15~$\mu$m and 22.5 mJy at 90~$\mu$m. 
Upper limits on the masses of circumpulsar dust are inferred at a given 
temperature using a simple modelling of the radiated flux; they are compared 
to upper limits of orbiting mass obtained with the dust heating model of Foster \& Fisher
 (1996). These results suggest that it is unlikely that any of these pulsars have 
sufficiently massive, circumpulsar discs, out of which planets 
may form in the future. 
}

\maketitle

\section{Introduction}
\label{sec:intro}
Pulsars are believed to be neutron stars born in supernova explosions, 
in which they receive a large kick due to an asymmetry in the 
explosion with a typical velocity $\ga 200\,$km/s.  Although they may 
be created with rapid spin rates, pulsars are thought to spin down 
rapidly because of their large initial magnetic fields ($\sim 
10^{12}\,$--$\,10^{13}\,$G) and corresponding short spin-down 
timescales (Lyne \& Graham-Smith 1998).  However, there is a class of pulsars 
characterized by a peculiar combination of very rapid (millisecond) 
spin rates and weak magnetic fields ($\sim 10^{8}\,$G). These objects 
occur preferentially in binary systems, with a binary fraction 
$\ga 50\,$percent, compared with $\sim 4\,$percent in the radio pulsar 
population as a whole (Lyne \& Graham-Smith 1998).  Such
pulsars are believed to originate from neutron stars born in binary 
systems which are spun up by accretion from their companion stars.\\  
This binary ``recycling'' model 
predicts that the companions of these pulsars must already be highly 
evolved objects near the end of their evolution, most likely white 
dwarfs or sometimes neutron stars, except in the case of systems where 
the companion is being evaporated (e.g., PSR 1957+20; Fruchter, Stinebring
\& Taylor 1988). The evaporation of the companion is due to heating by pulsar 
radiation, which may include electron-positron pairs and gamma-rays 
(Ruderman, Shaham \& Tavani 1989). 
This process can be understood as the final evolution 
of close low-mass X-ray binaries (see Bhattacharya \& van den Heuvel 
1991 for a detailed review). If the companion is evaporated completely, a 
single rapidly rotating recycled pulsar will remain. 
The discovery of at least three planet-mass objects orbiting the 
nearby millisecond pulsar B1257+12 almost a decade ago was a major 
surprise (Wolszczan \& Frail 1992). Indeed, this was the first 
planetary system discovered outside the solar system.  It is very 
unlikely that these planets existed around the progenitor of the 
pulsar and survived the supernova explosion in which the neutron star 
formed.  Therefore most pulsar-planet formation models postulate 
that the planets formed from a circumpulsar disc after the supernova 
explosion. The origin of these planet-forming discs is not well 
understood at the present time; there are numerous models in which the 
discs differ in their composition and physical properties 
(see Podsiadlowski 1993 and Phinney \& Hansen 1993 
for reviews and references). The discs could originate from fallback of 
supernova material, be surviving discs around massive stars 
or remnants of an evaporated companion. In the perhaps most promising 
class of models, the discs form out of the material of a companion star 
that was destroyed either as a result of a dynamical instability 
or in the supernova that formed the neutron star (because of a kick 
in the direction of the companion). In all of these latter models, one 
expects a disc of substantial mass (from a few tenths to a few \msun). 
Depending on whether the destroyed companion star was a normal-type 
star or a degenerate object (e.g., a CO white dwarf), the composition 
of the disc can range from solar-type material to a mixture dominated
by heavy elements. While some of these models require a millisecond pulsar, 
others do not and predict that planet-forming discs may exist around both 
recycled millisecond pulsars and normal radio pulsars. The initial conditions 
in a pulsar disc are probably extreme compared to normal protostellar nebulae; 
but as the disc expands and cools and the pulsar luminosity decreases, it 
may approach conditions more typical of discs around 
pre-main sequence stars (Phinney \& Hansen 1993; Ruden 1993). Indeed 
in some pulsar-planet formation models, the planet formation process 
itself could be very similar to the formation of our own solar system.\\ 
Searches for circumstellar material around neutron stars have been 
conducted for a handful of objects only, with a limiting sensitivity 
of $\sim 30\,$mJy at 10 \mic\} for warm ($T > 300\,$K) dust and at 
a limiting sensitivity of $\sim 10\,$mJy at sub-mm 
wavelengths for very cold ($T < 
30\,$K) dust, and none of them have shown any evidence for a 
circumpulsar disc. A sensitive search for 10 \mic\  continuum 
emission from PSR {B1257$+$12 has resulted in an upper limit of 
$7\,\pm\,11\,$mJy (Zuckerman 1993). Assuming that the circumstellar 
dust is cold ($T < 30\,$K), as might be expected if the pulsar 
spin-down luminosity is small or if the disc heating efficiency is 
low, Phillips \& Chandler (1994) searched for emission around five 
neutron stars in the sub-millimeter region (99 and 380 MHz). None of 
the pulsars in this sample was detected. Assuming that the circumpulsar 
discs were similar to those around T Tauri stars, they derived upper 
limits to the disc mass of $\sim 10^{-2}$\msun.\\ 
The Infra-red Space Observatory (ISO) with the spectro-photometer ISOPHOT 
was ideally suited to achieve high sensitivity in the intermediate 
temperature range, $30 < T < 300\,$K. In addition, ISOCAM 
in the range 12$-$18 \mic, allowed a search for warm dust of higher 
sensitivity than is possible from the ground. The main purpose of our study 
was 
to find evidence for circumpulsar discs, which might help to distinguish 
between different models for the origin of pulsar planets. In particular, 
we aimed to:\\ 
1) search for thermal dust emission from circumstellar discs or clouds 
(by-products or progenitors of the planet-formation process) around pulsars,\\ 
2) discover intermediate stages of evolution between evaporating binary 
pulsars and isolated millisecond pulsars with planets,\\ 
3) discover residual material from the envelope of the progenitor, that 
was not ejected in the supernova explosion and has settled in a post-supernova 
disc.\\
We also aimed to deduce the mass of radiating dust and compare its physical 
properties to that of dust in discs or shells around main-sequence and post 
main-sequence 
stars revealed by IRAS, ISO and ground-based infra-red and millimeter 
observations (see, e.g. Spangler et al.\ 2001).\\

\section{Observations and data reduction}
Our selected sample contains the nearest available pulsars known prior 
to August 1994: 3 millisecond pulsars and 3 ordinary radio pulsars, whose 
characteristics are shown on Table 1; note that
the nearby pulsar B1257+12 was not available because it was included in a guaranteed 
time programme with ISOPHOT and in a guest observer programme with ISOCAM; it is
 added to the Table for a later comparison (see 2.3).

\hspace*{1em} 
\begin{table} 
\centering 
\caption{The six pulsars observed with ISO (the pulsar B1257+12 is added for 
comparison)} 
\renewcommand{\arraystretch}{1.0} 
\setlength\tabcolsep{1.5pt} 
\begin{tabular}{lllllll} 
\hline\noalign{\smallskip} 
Pulsar & P & d  & log $ \dot E $  & companions & Reference  \\  
 & (s) & (pc) &(erg/s) & & \\
\hline 
\noalign{\smallskip} 
B1534$+$12 & 0.038 & 1080 & 33.25 & neutron star & Stairs et al 1998  \\ 
J2322$+$2057 & 0.0048 & 780 & 33.40 & isolated & Nice et al 1993  \\ 
J2019$+$2425 & 0.0039 & 910 & 33.73 &  white dwarf & Nice et al 1993  \\ 
B0149$-$16 & 0.8 & 790 & 31.95 & isolated & Siegman et al 1993  \\ 
B1604$-$00 & 0.42 & 590 &  32.21 & isolated & Philipps \& \\
 & & & & &Wolszczan 1992   \\ 
J0108$-$1431 & 0.85 & 85 & 30.78 & isolated & Tauris et al 1994  \\ 
\noalign{\smallskip} 
\hline 
\noalign{\smallskip} 
B1257$+$12 &  0.0062 & 620 & 34.30 & planets & Wolszczan (1993)   \\
\hline 
\end{tabular} 
\end{table}
\hspace*{1em}

We observed in the mid infra-red (MIR) at 15 \mic\ with ISOCAM 
(Cesarsky et al.\ 1996) 
and in the far infra-red (FIR) at 90 \mic\ with ISOPHOT (Lemke et al.\ 1996); 
each pulsar being observed during 1223 sec and 781 sec respectively, between 
1996, March and 1997, December. Preliminary results were presented in
Koch-Miramond et al, 1999. \\ 

\subsection{MIR observations and derivation of ISOCAM upper limits } 

\hspace{1em}
    Our additional motivation for the ISOCAM observation was to provide 
spatial resolution to a possible emission feature. The LW3 filter centered at 
15 \mic\ was used with a spatial resolution of 6 arcsec per pixel. 
The ISOCAM data were reduced with CIA version 3.0, following the
standard processing outlined in Starck et al.\ (1999). 
Transient corrections, using the inversion algorithm of 
Abergel, Bernard \& Boulanger (1996),  were applied. No 
detections were obtained at any of the pulsar positions.\\
Since the resulting maps gave no indication for infrared sources at
the expected source positions, we computed 3$\sigma$ upper limits in
the following way: (1) as there were no extended mid-infrared sources,
we computed the standard deviations of the noise present in the
maps. (2) We then assumed for each source that a point source remained
statistically insignificant while its peak had an amplitude less than
3$\sigma$. (3) We then used the known PSF profile to compute the total
source flux from the PSF peak value. In that last step, we had to make
another assumption, namely the location of the source inside the
ISOCAM pixel. Indeed, as ISOCAM generally undersamples the
instrumental PSF, the exact position of the source inside the pixel
can have a visible impact on the amount of light that falls in the
most illuminated pixel of the PSF. We assumed that the source fell at
the center of the pixel, which results in maximum light concentration.
A point source, brighter than our 3$\sigma$ upper limit, but falling
at the edge of a pixel, could still have its most illuminated pixel
fainter than 3$\sigma$. However, this configuration would result in
typically 2-4 equivalently bright pixels at the source location, which
we do not see in the maps. The derived 3$\sigma$ upper limits are between 
53 and 82 mJy,  (see Table 2). \\

\subsection{FIR observations with ISOPHOT}
\hspace{1em}
    We obtained ISOPHOT maps at 90 \mic\ at the positions of the six
pulsars using the oversampling mapping mode (AOT P32); the fields were
5 arcmin $\times$ 8 arcmin, with a 46 arcsec square aperture moved in
raster steps of 15 arcsec $\times$ 23 arcsec. The data
were reduced with version 6.1 of the PHT Interactive Analysis tool 
(PIA\footnote
{PIA is a joint development by the ESA Astrophysics Division and the ISOPHOT
consortium led by the Max--Planck--Institut f\"ur Astronomie, Heidelberg.}). 
No flux
enhancements were found at the radio positions of the pulsars except
for J0108$-$1431, the nearest known pulsar (Tauris et al.\ 1994), where a 
faint enhancement was observed. We therefore reduced the
field of J0108$-$1431 again with version 9.0 of PIA, using several
algorithms but no significant flux enhancement was obtained.\\ 
To derive upper limit for the 90\mic\ emission at the radio position of 
the pulsars, we measured the $\sigma$ values of the mean flux levels per detector 
pixel 46 arcsec square in a smooth mapped region around the pulsar position, 
after correcting for signal losses in the detector due to transients. 
From these measurements we derived 3 $\sigma$ upper limits 
between 22.5 and 130 mJy, (see Table 2).\\

\subsection{Upper limits on flux densities at 15 and 90 \mic\ from ISO  and comparison 
with the IRAS survey results}
\hspace*{1em}
    The 3 $\sigma$ upper limits on the ISO flux densities at 
15 and 90 \mic\ at the radio positions of the six pulsars 
are shown in Table 2.  Lazio et al (2001) report
60 and 90~$\mu$m observations of 7 millisecond pulsars (including J2322+2057) 
with ISOPHOT; their typical 3 $\sigma$ upper limits are 150 mJy.\\  
In view of the gain in sensitivity of about a factor 5 of the Scanpi  
\footnote{IRAS/Scanpi is a development made at IPAC/Infrared Science Archive, 
which is operated by the Jet Propulsion Laboratory, 
California Institute of Technology, under contract with the National
 Aeronautics and Space Administration.} 
processing of 
the IRAS survey over the IRAS Point Source Catalog (which has been used by van Buren \& Tereby 
(1993) to search for IRAS sources near the positions of pulsars), we considered the results 
of the Scanpi processing of all the IRAS scans 
passing within approximately 1.7 arcmin of the pulsar's positions. We carefully examined 
the coadded data; in most cases only upper limits can be defined; in a few cases 
a flux density deduced from the best-fitting point source template was detected at 
more than 2 $\sigma$ within the 1 arcmin beam of IRAS.  
 Both the 3 $\sigma$ flux limits and flux densities at 12, 25, 60 and 100~\mic\ are given  
in Table 2. Although these upper limits are not as stringent as the ISO ones, they put 
additional constraints on the derivation of the upper limits of circumpulsar masses. \\
The pulsar B1257$+$12 was added to our sample of six pulsars, not only 
for its intrinsic interest as the only known pulsar with planets
 but also because, together with B1534$+$12, it has published upper limits of fluxes in the 
mm and sub-mm ranges, which best constrain the upper limits on 
circumpulsar masses at low temperatures. At 850 \mic\, using the SCUBA instrument at JCMT, 
Greaves \& Holland (2000) 
obtained 3 $\sigma$ upper limits on the flux density of respectively 6.5 and 6.8 mJy, 
for B1534$+$12 and B1257$+$12; at 3.03 mm with the Owens Valley array Phillips \& Chandler 
(1994) obtained 3 $\sigma$ flux limits of 21 mJy for both pulsars. These two pulsars 
have also been observed at 10 \mic\ with the NASA Infra Red Telescope Facility by 
Foster \& Fischer (1996); they obtained 3 $\sigma$ upper limits on the flux density 
of respectively 32 and 27 mJy.} 




\hspace*{1em} 
\begin{table} 
\centering 
\caption{3 $\sigma$ upper limits on flux densities F in mJy from ISO at 15 and 90 \mic\ 
and 3 $\sigma$ upper limits on flux densities I in mJy from IRAS/Scanpi at 12, 25, 60 and 100 \mic; 
when there is a detection the 1 $\sigma$ error is given. } 
\renewcommand{\arraystretch}{1.5} 
\setlength\tabcolsep{2.0pt} 
\begin{tabular}{lllllll} 
\hline\noalign{\smallskip} 
Pulsar  & $\rm F_{\rm 15}$ & $\rm F_{\rm 90}$  & $\rm I_{\rm 12}$ & $\rm I_{\rm 25}$  & $\rm I_{\rm 60}$ 
 & $\rm I_{\rm 100}$  \\ 
\hline 
\noalign{\smallskip} 
B1534$+$12 & $<$82.2 & $<$75.0 &  $<$90 &  $<$90 & $<$140 & $<$390 \\   
J2322$+$2057  & $<$58.8 & $<$72.0 &  $<$100  & $<$110  & $<$120  &  $<$1200 \\ 
J2019$+$2425 & $<$64.5 & $<$130.0 &  $<$70  &  90$ \pm$30   & 140$\pm$ 60  &  $<$2100 \\ 
B0149$-$16 & $<$52.8 & $<$75.0 &  $<$110  &  $<$140  & 130$\pm$40  &  $<$300  \\ 
B1604$-$00 & $<$60.0 & $<$90.0 &  $<$90  &  $<$100 &  $<$120 & $<$480   \\ 
J0108$-$1431 & $<$66.0 & $<$22.5 & 170$\pm$40  &  $<$110  &  $<$90 &  250$\pm$124    \\ 
\noalign{\smallskip} 
\hline 
\noalign{\smallskip}

B1257$+$12 & & &  $<$130  & 200$ \pm$65  &  $<$120  &  $<$525  \\
\hline 
\end{tabular} 
\end{table} 
\hspace*{1em}

\section{Upper limits on circumpulsar masses}

We used a simple model to derive upper limits for the amount of circumstellar 
material in the form of grains. In the absence of indications on the dust 
composition provided by an accurate infrared spectrum of the dust, we
assumed that the dust is composed of interstellar grains
as described by Draine \& Lee (1984). From the optical constants 
for this material (a mixture of silicates and graphite with a ratio of 
$\sim$ 1.1 by particle number), one computes the mean absorption coefficients 
$Q_{\rm abs}$ of spherical particles as a function of the wavelength and the 
particle size using calculations based on the Mie theory (Bohren \& Huffman 1983).
Using the standard collisional size distribution, i.e. 
$n(a)da = A\,a^{-3.5}$ (Mathis, Rumpl \& Nordsiek 1977), 
where the constant $A$ ensures
the proper normalization of the distribution, the flux radiated
by a set of N particles at temperature $T_g$ can be written as
\begin{equation}
 F_{\nu}(\lambda) = N \int_{a_{\rm min}}^{a_{\rm max}} 4 \pi a^2\, 
Q_{\rm abs}(\lambda,a) \pi B_{\nu}
(\lambda,T_g)\, n(a)\, da
\end{equation} 
where $a_{min}$ and $a_{max}$ are the minimum and maximum sizes of the grains set to 0.01 
$\mu$m and 1000 $\mu$m, respectively, B$_{\nu}$ is the 
Planck function for blackbody emission per unit frequency. 
A lower cut-off size of 0.01 $\mu$m corresponds to the minimum size considered in
dust emission models by Lazio, Foster \& Fischer (2001); this minimum size is also
comparable to the minimum grain size inferred for the interstellar medium dust grains
(Mathis \& Whiffen, 1989). The maximum size is arbitrarily fixed to 1 mm, a size
above which the integrated emission of the dust over the wavelength of interest 
(roughly 5 $\mu$m to 3 mm) is 10$^{-13}$ times lower than the integrated 
emission of the particles with sizes in the range 0.01 to 1000 $\mu$m (which means
that we have currently no constraints on the mass of particles bigger than 1 mm). 
The influence of the minimum cut-off size is studied in Figure 1 in which upper limits
on the dust mass are plotted for the case of PSR B1534$+$12, 
 parametrized by the minimum cut-off size.\\

\begin{figure}[!hp] 
\includegraphics[width=8.5cm]{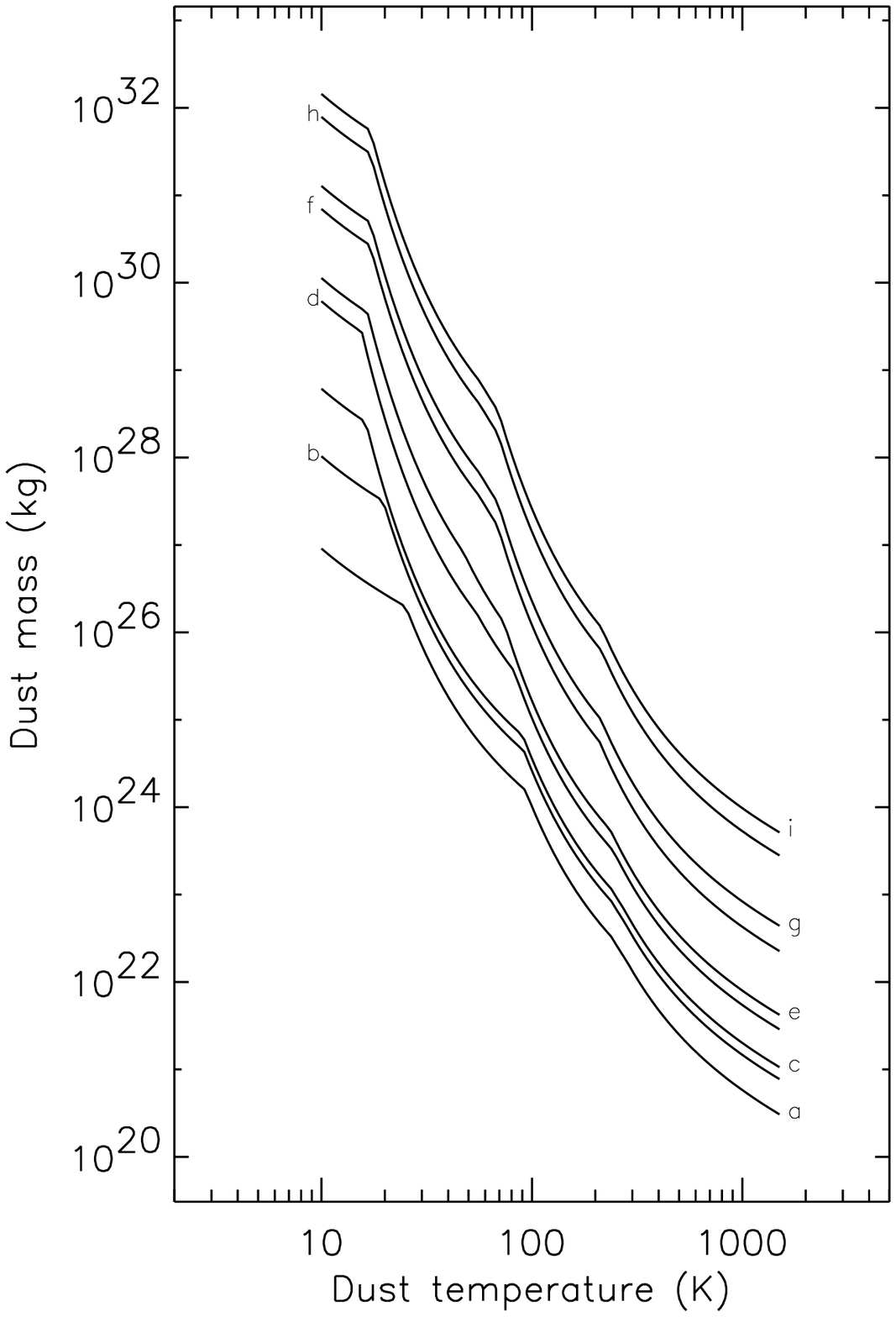}
\caption{Influence of the minimum cut-off in the grain size distribution on the 
upper limits of dust masses derived for PSR B1534$+$12 in the temperature range of
10 - 1500 K; cut-off values: a = 0.01 \mic, b = 0.05 \mic, c = 0.1 \mic, d = 0.5 \mic, 
 e = 1 \mic, f = 5 \mic, g = 10 \mic, h = 50 \mic, i = 100 \mic.   
} 
\label{Fig1}
 
\end{figure}

The range of circumpulsar mass limits allowed in the above model by our ISO data 
and the IRAS/Scanpi data, 
and for B1534+12 and B1257+12 by the published sub-mm and mm data are shown on 
Figure 2.

 

Each point in the plots of Figure 2 represents an upper limit on the mass
for a given temperature, the range of temperatures being chosen between 10 K (typical lower
temperature of interstellar cold dust) and 1500 K (sublimation temperature
of silicate dust). For each temperature, a probability density (coded by a grey-level on the left 
bars in the plots, the color of the points being reported on the bar) is computed
by combining the partial probability density functions for each data point. 
A data point with a true value is assumed to follow a 
Gaussian partial probability density function with a standard deviation deduced from
the error on each data point. A data point which corresponds to a lower limit is assumed
to follow a half Gaussian-like partial probability density function for values
greater than the data point value and an uniform probability density function
for lower values. Most probable values for the temperatures are shown in the plots 
as the brightest points. The error bars overplotted correspond for each temperature point to 
1/1000 of the maximum density of probability. Although the detections obtained in the 
IRAS beam at the position of pulsars are probably chance coincidences (van Buren \& Tereby 1993), 
their influence on the most probable temperature of the grains is clearly seen in Figure 2. 
The extremas of dust mass upper limits corresponding to temperatures 10 K and 1500 K are shown in Table 3.\\

We have also tested the global dust heating model used by Foster \& Fischer (1996)
which assumes that a fraction of the pulsar's spin-down
luminosity is heating a dust disc and gives a relation between the total dust mass in the disc 
and the temperature. This
dependance is shown in Figure 2 as a dashed line overplotted. The pulsars' spin-down
luminosities are shown in Table 1; for each pulsar, 
the parameter f expressing the fraction of spin-down
luminosity converted into dust thermal energy is taken as 1 percent (Foster \& Fischer, 1996). 
Figure 2 shows that there is a temperature T$_{\rm cr}$ corresponding to the same upper limit 
of circumpulsar mass in the two models, if we allow f to increase slightly above 1 percent. 
These temperatures T$_{\rm cr}$ are shown in Table 3 
together with the corresponding upper limits of circumpulsar masses M$_{\rm cr}$ in solar mass units. 
     
\hspace*{1em}
 
\begin{table} 
\centering 
\caption{Upper limits on mass of emitting dust around pulsars computed at temperatures 
T$_g$ = 10 K and 1500 K; and upper limits on mass at a temperature T$_{\rm cr}$ deduced from the model of 
Foster \& Fischer (1996)  
} 
\renewcommand{\arraystretch}{1.5} 
\setlength\tabcolsep{5pt} 
\begin{tabular}{llllll} 
\hline\noalign{\smallskip} 
Pulsar & M$_{\rm 10K}$   & M$_{\rm 1500K}$ & T$_{\rm cr}$   & M$_{\rm cr}$ 
 &  M$_ {\rm cr}$/${\mathrm M}_\odot$  \\ 
 &(kg) & (kg) & (K)  & (kg)  & \\
\noalign{\smallskip} 
\hline 
\noalign{\smallskip} 
B1534$+$12   & $<$$10^{27}$ & $<$$10^{20}$ & 10 & $<$$10^{27}$ & $<$$5\;10^{-4}$ \\ 
J2322$+$2057 & $<$$10^{30}$ & $<$$10^{20}$  & 30 & $<$$10^{26}$ & $<$$5\;10^{-5}$  \\ 
J2019$+$2425 & $<$$2 10^{30}$ & $<$$10^{20}$  & 30 & $<$$10^{26}$ & $<$$5\;10^{-5}$ \\ 
B0149$-$16   & $<$$10^{30}$ & $<$$10^{20}$ & 30 & $<$$10^{25}$ & $<$$5\;10^{-6}$ \\ 
B1604$-$00   & $<$$10^{30}$ & $<$$10^{20}$ & 30 & $<$$10^{25}$ & $<$$5\;10^{-6}$ \\ 
J0108$-$1431 & $<$$10^{28}$ & $<$$2\;10^{18}$  & 30 & $<$$10^{23}$ & $<$$5\;10^{-8}$ \\ 
\noalign{\smallskip} 
\hline 
\noalign{\smallskip}

B1257$+$12 & $<$$10^{24}$  & $<$$10^{19}$ & 60 & $<$$10^{23}$ & $<$$5\;10^{-8}$  \\
\hline 
\end{tabular} 
\end{table}
\hspace*{1em}

We note that the latter upper limit of circumpulsar 
mass for PSR B1534$+$12 is 
30 times smaller than the upper limit of $1.6\times 
10^{-2}$\msun\ obtained by Phillips \& Chandler (1994) in the sub-mm 
and mm ranges, using the Beckwith et al (1990) results on 
circumstellar discs around T Tauri stars. Greaves \& Holland (2000) using 
their upper limits of flux at 850~$\mu$m for
B1534$+$12 and B1257$+$12, and the Foster \& Fisher (1996) model 
with grain size 100 \mic\ and a spin-down luminosity set at $2\times 10^{34}$ erg/sec 
for both pulsars,
deduced upper limits to disc masses typically 
lower than 10 Earth masses i.e. $<$ $3\times 10^{-5}$ ${\mathrm M}_\odot$.   
   
\section{Discussion and Conclusions}

These upper limits for the dust mass around pulsars M$_{\rm cr}$/${\mathrm M}_\odot$, 
suggest that none of them are surrounded by a sufficiently 
massive disc in which planets are likely to form. It is generally agreed that 
the suitable protoplanetary 
disc has at least 0.01 ${\mathrm M}_\odot$ of gas and dust in Keplerian 
orbit around a solar-mass protostar (Boss 2000). 
The dust mass found in T Tauri discs is typically
 $10^{-3}{\mathrm M}_\odot$ (Beckwith et al 1990); 
 protoplanetary discs with masses in the range of 0.01 to 0.1 ${\mathrm M}_\odot$ 
are commonly found in orbit around young stars (Zuckerman 2001). When stars reach ages of about
$10^{7}$ yr, the evidence of planet-forming discs disappears (Boss 2000). Evidence of
postplanetary discs has been found, f.i. around Beta Pictoris  with about
$10^{-6}$${\mathrm M}_\odot$ (Artymowicz 1994). 
One caveat is that these estimates are very dependent on the properties of the 
dust grains and do not provide a good estimate for the total amount of 
gas and dust because the dust to gas ratio is undetermined. These estimates 
could be quite different for circumpulsar discs with very non-solar 
composition. This negative result is perhaps 
not so surprising, since planets 
around pulsars do not appear to be common observationally 
(Konacki, Maciejewski \& Wolszczan 1999), 
certainly much rarer than planets around normal-type stars. This also 
suggests that planet formation around pulsars is not a natural 
consequence of the pulsar-formation process (whether it is the 
formation of the neutron star in a supernova or the recycling of the 
pulsar in a binary). This is rather different from planet 
formation around normal-type stars, which appears to be an ubiquitous 
by-product of the star-formation process. 

\begin{acknowledgements}
~ \\
We warmly thank the ISO project and the ISOCAM and ISOPHOT Teams in
Villafranca, Saclay and Heidelberg. This research has made use of the NASA/ IPAC 
Infrared Science Archive, which is operated by the Jet Propulsion Laboratory, 
California Institute of Technology, under contract with the National
 Aeronautics and Space Administration. We express our thanks to the anonymous 
referee for very helpful comments. 

\vspace*{2ex}

\end{acknowledgements}

\end{document}